\documentclass[11pt]{nih2}

\usepackage[pdftex]{graphicx}
\usepackage{fancyhdr}

\begin{document}

\begin{center}
\Large\textbf{Electroporation dynamics for different pore lifetimes \\
based on the standard model} \\
\ \\
\normalsize
Thiruvallur R. Gowrishankar,$^1$
Julie V. Stern,$^1$ and
James C. Weaver$^{1,2}$\\
$^1$Harvard-MIT Division of Health Sciences and Technology,\\
Massachusetts Institute of Technology, Cambridge, MA 02139, USA\\
$^2$Corresponding author: jcw@mit.edu\\
\end{center}

\section*{Abstract} 

Standard model of electropration (EP) has long emphasized a single
pore lifetime to explain post-pulse transport across cell membranes.
However, pore lifetimes estimated from molecular dynamics (MD) models and 
and those measured from experimental data differ by several orders of
magnitude.  We hypothesize that a broad distribution of lifetimes may
describe the post-pulse behavior.  Here, we show that pore distribution,
number and size of pores, show interesting behavior in different ranges
of pore lifetimes.   Interestingly, for large electric fields ($>$ 1 kV/cm) 
and short pore lifetimes ($\sim$100 ns), an siginificant loss in pore 
number occurs during the pulse.   Given the large number of EP applications
that apply such fields, this phenomenon may be crucial to post-pulse 
response of cell membrane to such fields.

\section{Introduction} 

Lipid bilayers that constitute cell membranes maintain ionic and moelcular
concentration gradient across the membranes.  
Electroporation (EP), a non-thermal response of lipidic membranes 
to an electric field pulse, compromise the barrier to 
transmembrane transport of ions and membranes.  
The highly nonlinear and hysteretic phenomenon
of the standard EP model results from the creation, 
evolution and destruction of membrane pores
\cite{KrassowskaFilev_ModelingElectroporationSingleCellSphericalPoreExpansion_BPJ2006,%
EsserEtAl_IntracellularManipulationConventionalEP_BPJ2010,%
TaleleGaynorEtAl_ModelSingleCellElectroporationDynamicPoresBipolarPulses_JElectrostat2010,%
SmithKC_CellModelWithDyamicPoresAndElectrodiffusionofChargedSpecies_DoctorateThesisMIT_2011,%
WeaverChizmadzhevChapterCRC1996,%
Weaver_Review_SpecialIssue_IEEE_TransactionsDielectricsElectricalInsulation2003}.
Typical EP fields range in amplitude from a few hundred V/cm to hundreds 
of kV/cm and in duration from a few nanoseconds to hundreds of 
milliseconds 
\cite{WeaverEtAl_StrengthDuration_Bchem2012}.

Cell EP is already well established in medicine
\cite{YarmushEtAl_EP_BasedTechnologiesForMedicine_AnnuRevBiomedEng2014}, 
specificially in applications such as in vitro and
in vivo gene transfection
\cite{GothelfAndGehl_EP_Based_DNA_Vaccines_HumVaccinImmunother2012}, 
delivery of bioactive molecules by EP 
\cite{VenslauskasEtAl_SmallMoleculeDeliveryEP_Bioelectrochem2010}, and
analysis of intracellular contents and delivery of small molecules by 
microfluidic EP 
\cite{GengAndLu_MicrofluidicEP_LabChip2013}.
In addition, four distinct methods of tissue ablation based on EP are 
being pursued, {\it viz.,} electrochemotherapy (ECT) 
\cite{CalvetAndMir_ECTAndImmunotherapy_CancerMetastRev2016, 
BretonMir_MicrosecondNanosecondElectricPulsesCancerTreatments_BEMS2012},
irreversible electroporation (IRE)
\cite{JiangEtal_IRE_Review_TBME2015}, 
calcium-EP \cite{FrandsenEtal_NormalAndMalignantCellsResponsesCalciumEP_CancerRes2017}
and EP by  nanosecond pulsed electric field (nsPEF) pulses
\cite{Beebe_nsPEFMechanisms_JNanomedRes2015,%
ChopinetAndRols_nsPEF_Review_Bioelectrochem2015}. 
Recently, an emerging application based on bipolar cancellation of 
cellular responses to nanoelectroporation has gained attention
\cite{PakhomovEtAlIbey_CancellationCellularResponses-nsPEF-StimulusPolarityReversal_CellMolLifeSci2014,%
GianulisEtAlPakhomov_ElectroporationMammalianCellsNanosecondFieldOscillations-FieldReversalInihibition_SciReports2015}.

All aspects of pore dynamics are governed by the transmembrane 
voltage-dependent pore energy or
landscape, $\mathrm{W(r_p,U_m})$.  Pore creation and destruction are 
described by two absolute rate equations that depend on the pore energy
landscape   
\cite{SmithKC_CellModelWithDyamicPoresAndElectrodiffusionofChargedSpecies_DoctorateThesisMIT_2011,%
VasilkoskiEtAl_ElectroporationAbsouteRateEquationNanosecondPoreCreation_PhysRevE2006}.
Pores created with radius $\mathrm{r_p < r_*}$ are considered short-lived, 
and hydrophobic.   These pores are rapidly destroyed by thermal fluctuations
in the membrane
\cite{AbidorEtAlChizmadzhevMainFacts1stPaper1979}.
Pore expansion and contraction are modeled by the Smoluchowski
equation that incorporate the pore energy landscape. 
Cell membrane EP models determine the EP dynamics using the free energy
cost of creating membrane pores of a particular pore radius.  This energy,
$W$, is reduced by an elevated transmembrane voltage.   

EP experiments report pore lifetimes ($\mathrm{\tau_p}$)
ranging from $\sim$100~ns to  $\sim$1,000~s.
Traditionally these are based on measurements of post-pulse membrane
recovery, usually termed ``resealing" 
\cite{WeaverAndVernier_ComplexDarkPores_arXiv2017}.
Some complex events may occur as TPs and CPs are simultaneously formed, 
with molecules and molecular segments drawn into CPs as they form
\cite{SternEtAl_BroadPoreLifetimeDistributions_arXiv2017}.
Such structures may be fleeting, but may contribute to solute rectification, 
with net transport persisting long afterwards.
Molecular dynamics (MD) models of traditional lipidic pores have lifetimes,
$\tau_{\mathrm{p}}$, of order $100$~ns 
\cite{BennettEtAlTieleman_AtomisticSimulationsPoreFormationClosure_BLM_BPJ2014,%
LevineVernier_LifeCyclePoreStepsCreationAnnihilation_JMemBiol2010}.  
A longer
$\tau_{\mathrm{p}}$ lifetime estimate is 0.02~s,
based on recovery between pulses in a pulse train
\cite{PakhomovEtAlPakhomova_MultipleNanosecondPulsesIncreaseNumber-NotLongLivedPoreSize_BBA2015}.
Modeling 
of calcein uptake by prostate cancer cells
\cite{CanatellaEtAlPrausnitz_ElectroporationUptakeViabilityQuantitativeStudy_BPJ2001} determines a lifetime of $\sim$4~s 
\cite{SmithKC_CellModelWithDyamicPoresAndElectrodiffusionofChargedSpecies_DoctorateThesisMIT_2011}.
Another experiment 
\cite{PakhomovEtAl_nsPEF_LongLastingSinglePulsePermeabilizationPM_BEMS2007}
shows a long-lasting membrane permeability 
with a post-pulse recovery of PM extending to $\approx$ 180~s. 
In addition, other experiments 
\cite{KennedyEtAlBooske_QuantificationElectroporationKineticsPI_FinalVersion_BPJ2008,%
PakhomovaEtAlPakhomov_Electroporationn-InducedElectrosensitization_PLoS_ONE2011}
show a delayed uptake of propidium over $\sim$300~s after the pulse.
Recently, we hypothesized that complex pores in membrane lead to a broad
distribution of pore lifetimes
\cite{WeaverAndVernier_ComplexDarkPores_arXiv2017,%
SternEtAl_BroadPoreLifetimeDistributions_arXiv2017}.
Here, we look at the EP response of a 
cell membrane model for two pore lifetimes that differ by several orders
of magnitude.

\section{Methods} 

\subsection{Pore energy}

The dynamic behavior of pores is determined by the pore energy landscape
\cite{SmithKC_CellModelWithDyamicPoresAndElectrodiffusionofChargedSpecies_DoctorateThesisMIT_2011}.  The transmembrane voltage-dependent pore energy, 
$\mathrm{W}$, defines the creation, expansion, contraction and destruction
of pores.  
Pore energy is the sum of mechanical and electrical energy 
\begin{eqnarray*}
W(r_p,\Delta\phi_m) & = & W_{mech}(r_p) + W_{elec}(r_p,\Delta\phi_m),\\
\end{eqnarray*}
where
\begin{eqnarray*}
W_{mech}(r_p) & = & B\left(\frac{r_*}{r_p}\right)^b + C + 2\pi\gamma r_p - \Gamma \pi r_p^2\\
W_{elec}(r_p,\Delta\phi_m) & = & -F_{max}\left(r_p+r_h\ln\left(\frac{r_t+r_h}{r_p+r_t+r_h}\right)\right)(\Delta\phi_m)^2 \\
\end{eqnarray*}

The mechanical energy results from three contributions: steric repulsion 
energy, edge energy, and surface interfacial energy.  The elctrical energy is 
 energy from electrical force on expanding force. 

The resealing time constant $\tau_p$
is approximately related to the energy barrier to pore destruction $W_d$
by 
\cite{SmithKC_CellModelWithDyamicPoresAndElectrodiffusionofChargedSpecies_DoctorateThesisMIT_2011}
\begin{eqnarray*}
\tau_p \approx \frac{(r_m-r_*)^2}{D_p}\left(\frac{W_d}{kT}\right)^{-\frac{3}{2}} e^{W_d/kT}
\end{eqnarray*}
The equation above is solved numerically to find the value of the pore 
destruction barrier $W_d$ that results in approximately the desired resealing 
time constant $\tau_p$.  The algorithm adjusts the value of $W_d$ until
the resulting $\tau_p$ is equal to the desired value.

Pore energy landscape shows the behavior of pore energy as a function of 
pore radius for different transmembrane voltages.  
Different energy landscapes (Fig.~1) were generated with mostly 
identifical parameters
for pore lifetimes of 100 ns and 4 s.  The only parameter that was 
changed was the pore radius at which the pore energy minimum is located.\\

\subsection{Cell model}

We use a meshed cell membrane model 
\cite{SmithEtAl_EmergencelargePoreSubpopulationDuringElectroporationPulses2013,
SmithKC_CellModelWithDyamicPoresAndElectrodiffusionofChargedSpecies_DoctorateThesisMIT_2011}
to simulate the dependence of pore lifetime on the response of 
the model to different applied fields.
The model includes transmembrane voltage-driven pore creation,
and stochastic pore
destruction in a typical mammalian cell plasma membrane (PM).
We employ a dynamic local model of EP based on a meshed transport 
network system model 
\cite{SmithEtAl_EmergencelargePoreSubpopulationDuringElectroporationPulses2013,
SmithWeaver_ActiveMechanismsNeededSubmicrosecondMVperMeterPulses_BPJ2008,
GowrishankarEtAl_TransportBasedModelsElectricFieldCellResponses_Review_ProcIEEE2012}
consisting of a cylindrical cell membrane, intracellular and extracellular 
solutions, and idealized electrodes at the boundaries of the system model
\cite{SmithEtAl_EmergencelargePoreSubpopulationDuringElectroporationPulses2013,
SmithKC_CellModelWithDyamicPoresAndElectrodiffusionofChargedSpecies_DoctorateThesisMIT_2011}
(Fig.~2).
The plasma membrane (PM) is $5 \mathrm{\, \mu m}$ in 
radius, $13.3 \mathrm{\, \mu m}$ in length, and 4~nm thickness [see Fig.~2(a)].

Cell EP involves spatially distributed, highly nonlinear and hysteretic 
interactions throughout the system model,
and can only be solved computationally
\cite{DeBruinKrassowska_TheoreticalModel_SingleCellEporeI_FieldStrength_RestPotential_BPJ1999,%
KrassowskaFilev_ModelingElectroporationSingleCellSphericalPoreExpansion_BPJ2006,%
SmithWeaver_ActiveMechanismsNeededSubmicrosecondMVperMeterPulses_BPJ2008,%
JoshiSchoenbach_Review_nsPEF_CritRevBME2010,%
EsserEtAl_IntracellularManipulationConventionalEP_BPJ2010,%
LiLin_NumericalSimulationMolecularUptakeElectroporation_BChem2011}.
Features of the present model
\cite{SmithEtAl_EmergencelargePoreSubpopulationDuringElectroporationPulses2013, SmithKC_CellModelWithDyamicPoresAndElectrodiffusionofChargedSpecies_DoctorateThesisMIT_2011}
include mesh creation for a 150 node-pair plasma membrane and 9267 
nodes for the intra- and extracellular aqueous electrolytes, 
with charge transport and storage 
between mesh nodes [see Fig. 1(b)-(c)]. 
We also include a -50~mV resting potential source for each membrane site, 
which is shunted by increased membrane conductance due to pores.

The model geometry, meshing, and electroporation parameters are identical to 
the model used in recent work examining the dynamic pore population response to 
a range of electroporating pulses 
\cite{SmithEtAl_EmergencelargePoreSubpopulationDuringElectroporationPulses2013,
SmithKC_CellModelWithDyamicPoresAndElectrodiffusionofChargedSpecies_DoctorateThesisMIT_2011}. 
One important feature is a realistic and non-trivial computation of rapidly changing membrane conductance.
This is complicated because of non-linear individual pore conductance arising 
from both pore size and local transmembrane voltage, 
$\Delta\phi_m(t)$
\cite{DeBruinKrassowska_TheoreticalModel_SingleCellEporeI_FieldStrength_RestPotential_BPJ1999,%
KrassowskaFilev_ModelingElectroporationSingleCellSphericalPoreExpansion_BPJ2006,%
SmithKC_CellModelWithDyamicPoresAndElectrodiffusionofChargedSpecies_DoctorateThesisMIT_2011,%
SmithEtAl_EmergencelargePoreSubpopulationDuringElectroporationPulses2013}.

\subsection{Applied field}

Three different electric field pulses are employed (Fig.~3).  These are 
a 13.3 kV/cm, 60 ns experimental nsPEF pulse
\cite{PakhomovaEtAlPakhomov_Electroporationn-InducedElectrosensitization_PLoS_ONE2011},
an idealized 1.5 kV/cm, 100 $\mathrm{\mu}$s trapezoidal pulse with 
1 $\mathrm{\mu}$s rise and fall times, and a 0.8 kV/cm, 100 ms trapezoidal
pulse with 
1 $\mathrm{\mu}$s rise and fall times 
\cite{SadikEtAlLin_QuantiativePropidiumIodideExperimentsModel-ms-duration_BBA2013}.  Model parameters are listed in Table~1.

\section{Results} 

\subsection{Transmembrane voltage response}

The transmembrane voltage response shows different behavior during the pulse 
for the three different field pulses (Fig.~4).  The large magnitude (13.3 kV/cm)
nsPEF
pulse creates a large number of pores in the membrane causing a high
conductance shunt in the membrane.  This leads to the reduction of 
transmembrane voltage to $\sim$0.5 V (Fig.~4 left) for the rest of the pulse
(reversible electrical breakdown; REB).  Following the pulse, the 
transmembrane voltage decreases nearly to zero. The REB behavior is also
seen with the 1.5 kV/cm conventional EP pulse (Fig.~4 middle).
However, the amplitude of the long pulse (0.8 kV/cm) is not 
sufficiently large to create a large number of pores.  Consequently,
this pulse does not exhibit REB (Fig.~4 right).

\subsection{Total pore number}

The most interesting result is the decline in pore number, $\mathrm{N(t)}$,
during the pulse
for short pore lifetimes (100 ns; Fig.~5, top row).  When the field 
amplitude is large enough
to create over 1,000 pores, the membrane conductance increases by several
orders of magnitude causing the transmembrane voltage to drop during the
pulse (REB).  This prevents the pores from expanding and given the short
pore lifetime, a fraction of existing pores disappear during the pulse.
This phenomenon is striking for a 1.5 kV/cm, 100 $\mathrm{\mu}$s pulse 
(Fig.~5, top middle).  This effect is also seen for a nsPEF pulse 
(Fig.~5, top left) although not as pronounced as that seen with
a conventional pulse.  However, when the field strength is not large
enough to create a REB, as in the 0.8 kV/cm, 100 ms pulse, many fewer 
pores are created and the conductance is not large enough to cause 
a decline in pore number (Fig.~5, top right).

The decline in pore number is, however, not seen when the pore lifetime
is 4 s (Fig.~5, bottom row).  The pore lifetime in this case is much
longer than the pulse duration ensuring that the pores created at the
onset of the pulse do not decay before the pulse ends.  The maximum number 
of pores is nearly the same for both pore lifetimes for a given field
pulse.

\subsection{Pore dynamics}

Evolution of pore distribution for different field pulses and the
two pore lifetimes are shown in Figs.~6-8.  

A short duration, large amplitude (nsPEF; 13.3 kV/cm, 60 ns) 
pulse creates a large
number of pores (Fig.~6).  However, the short duration of the pulse 
precludes the pores from expanding much during the pulse.  At the
onset of the pulse (Fig.~6 left), there is a slight asymmetry between
anodic (top panel) and cathodic (bottom panel) sides because of the
resting transmebrane voltage (-50 mV).  Most of 
the pores remain close to 0.7 nm in radius. In the case of 100 ns pore
lifetime (Fig.~6, top row), most of the 
pores disappear shortly after the pulse ends, the number of pores decreases
from nearly 400,000 at the end of the pulse to 100 within 1 $\mathrm{\mu}$s
(Fig.~6, top center left and top center right).   By 1 s, the pore distribution
relax to a thermalized distribution (Fig.~6, top right).  
In contrast, when the pore lifetime
is 4 s, most pores remain in the membrane even at 1 s after the pulse ends
(Fig.~6, bottom right).  In the 4 s case, the number of pores decreases
from 400,000 at the end of the pulse to nearly 300,000 at 1 s post-pulse.

In the case of a conventional EP pulse (1.5 kV/cm, 100 $\mathrm{\mu}$s
trapezoidal pulse), the total number of pores is much smaller
smaller than from a nsPEF pulse.   However, the pores begin to expand
even as early as 1 $\mathrm{\mu}$s into the pulse.  Most pores expand
to 12 nm (maximum pore size in the model) thus increasing the membrane
conductance and preventing new pores being created (Fig.~7, top left and 
top center left) in the case of 100 ns pore lifetime.    Once the pulse
ends, the pores relax back toward the minimum pore radius.  By 1 s 
after the pulse, the pore distribution has relaxed to a thermal 
distribution in the case of a 100 ns pore lifetime (Fig.~7, top right).
When the pore lifetime is increased to 4 s, pore creation competes with
pore expansion leading to nearly equal number of pores at 
$\mathrm{r_{min}}$ and $\mathrm{r_{max}}$.  Following the pulse, the
number of pores relaxes much more slowly (with a time constant of 4 s).
As seen in Fig.~7, bottom right, over 600 pores still remain 1 s post-pulse 
on both anodic and cathodic sides.

A much longer, but lower amplitude pulse (100 ms, 0.8 kV/cm) creates 
even less number of pores in the membrane (Fig.~8).  But, all the pore expand
to the maximum pore radius by the end of the pulse (Fig.~8, top center left).
Once the pulse ends, the pore distribution begins to contract toward a 
thermalized distribution.   By 1~s post pulse, only the minimum-sized
pores remain in the membrane (Fig.~8, right) for both 100 ns and 4 s
pore lifetimes.  However, most of the pores have vanished in the short
pore lifetime cases while a significant number of pores remain in the
case of 4 s pore lifetime.

\section{Discussion}

Pore lifetime is a basic measure of pore destruction.  It is often
determined from exponentially decaying membrane conductance or permeability.
The traditional view of a single pore lifetime does not adequately represent
the complex behavior of a cell membrane response during and after an 
electric field is applied.  The dynamics of membrane response to such
fields is complicated by the presence of a number of biological 
molecules in different conformational states and binding affinities.

Our results suggest that a fraction of the pores with very short lifetimes
(shorter than the pulse duration) may disappear during the pulse if their
numbers are large enough to cause a reverisble electrical breakdown (REB)
of the membrane.  However, if the pulse is not sufficiently large, it may
create fewer pores that expand during the pulse without causing an overall
decrease in the pore number.  These differences in post-pulse pore 
distribution (size and number) may lead to significant differences in
the transmembrane transport of various size molecules.  This has a broad
impact on many EP applications that employ fields that span several orders
of magnitude in duration and strength.

The long EP pulse (0.8 kV/cm, 100 ms) shows a steady decline in 
transmembrane voltage during the pulse.  Unlike most optical measurements, 
electrical response of the
cell membrane during the pulse could be measured with sufficient time
resolution to investigate the decline in transmembrane voltage during 
the pulse.  A long, low amplitude pulse does not create a large number
of pores, but as the pores expand, the membrane conductance increases
leading to a decrease in transmembrane voltage.  This behavior could 
be verified using field pulses in the range of a few hundred kV/cm and
durations of hundreds of milliseconds.

Recent experiments using optical single-channel recording 
provide insight into EP dynamics
\cite{SengelWallace_ImagingDynamics-IndividualElectropores_PlanarDropletInterfaceBilayers_PNAS2016,%
SengelAndWallace_PotentialEnergyBarrier_BLM_EP_PhilTransRoySocB2017}.
These experiments confirm that EP behavior is not characterized by 
a single dynamic but by a range of fluctuation kinetics. In addition,
the experiments determine the energy barrier to pore opening to be
around 25 kT.  

Pore energy landscapes with different pore lifetimes help in exploring the
response of membrane at different time scales, during, immediatley after the
pulse and long after the pulse.  These descriptions are particularly useful
in explaining experiments that demonstrate post-pulse membrane transport
for hundreds of seconds to several minutes.  In addition, the pore energy
landsacpes may also assist in understanding the phenomenon 
of bipolar cancellation that has been reported in different cell types
\cite{PakhomovEtAlIbey_CancellationCellularResponses-nsPEF-StimulusPolarityReversal_CellMolLifeSci2014,%
GianulisEtAlPakhomov_ElectroporationMammalianCellsNanosecondFieldOscillations-FieldReversalInihibition_SciReports2015}.

\section*{Acknowledgements} 

This work was supported by AFOSR MURI grant FA9550-15-1-0517 on 
Nanoelectropulse-Induced
Electromechanical Signaling and Control of Biological Systems, 
administered through Old Dominion University.
We thank P. T. Vernier, and A. G. Pakhomov
for multiple stimulating discussions,
and K. G. Weaver for computer support.

\section*{References}
\def\refname{}
\bibliographystyle{unsrt}

\begin{thebibliography}{}

\end{thebibliography}


\begin{thebibliography}{10}

\bibitem{KrassowskaFilev_ModelingElectroporationSingleCellSphericalPoreExpansion_BPJ2006}
W.~Krassowska and P.~D. Filev.
\newblock Modeling electroporation in a single cell.
\newblock {\em Biophys. J.}, 92:404--417, 2007.

\bibitem{EsserEtAl_IntracellularManipulationConventionalEP_BPJ2010}
A.~T. Esser, K.~C. Smith, T.~R. Gowrishankar, Z.~Vasilkoski, and J.~C. Weaver.
\newblock Mechanisms for the intracellular manipulation of organelles by
  conventional electroporation.
\newblock {\em Biophys. J.}, 98:2506--2514, 2010.

\bibitem{TaleleGaynorEtAl_ModelSingleCellElectroporationDynamicPoresBipolarPulses_JElectrostat2010}
S.~Talele, P.~Gaynor, M.~J. Cree, and J.~van {E}keran.
\newblock Modelling single cell electroporation with bipolar pulse parameters
  and dynamic pore radii.
\newblock {\em J. Electrostatics}, 68:261--274, 2010.

\bibitem{SmithKC_CellModelWithDyamicPoresAndElectrodiffusionofChargedSpecies_DoctorateThesisMIT_2011}
K.~C. Smith.
\newblock {\em A unified model of electroporation and molecular transport}.
\newblock Massachusetts Institute of Technology,
  http://dspace.mit.edu/bitstream/handle/1721.1/63085/725958797.pdf.

\bibitem{WeaverChizmadzhevChapterCRC1996}
J.~C. Weaver and Y.~A. Chizmadzhev.
\newblock Electroporation.
\newblock In C.~Polk and E.~Postow, editors, {\em Handbook of Biological
  Effects of Electromagnetic Fields}, pages 247--274. CRC Press, Boca Raton,
  2nd edition, 1996.

\bibitem{Weaver_Review_SpecialIssue_IEEE_TransactionsDielectricsElectricalInsulation2003}
J.~C. Weaver.
\newblock Electroporation of biological membranes from multicellular to nano
  scales.
\newblock {\em {IEEE} Trans. Dielect. Elect. Ins.}, 10:754--768, 2003.

\bibitem{WeaverEtAl_StrengthDuration_Bchem2012}
J.~C. Weaver, K.~C. Smith, A.~T. Esser, R.~S. Son, and T.R. Gowrishankar.
\newblock A brief overview of electroporation pulse strength –- duration
  space: A region where additional intracellular effects are expected.
\newblock {\em Bioelectrochemistry}, 87:236--243, 2012.

\bibitem{YarmushEtAl_EP_BasedTechnologiesForMedicine_AnnuRevBiomedEng2014}
M.~L. Yarmush, A.~Goldberg, G.~Sersa, T.~Kotnik, and D.~Miklavcic.
\newblock Electroporation-based technologies for medicine: Principles,
  applications, and challenges.
\newblock {\em Annu. Rev. Biomed. Eng.}, 16:295--320, 2014.

\bibitem{GothelfAndGehl_EP_Based_DNA_Vaccines_HumVaccinImmunother2012}
A.~Gothelf, , and J.~Gehl.
\newblock What you always needed to know about electroporation based dna
  vaccines.
\newblock {\em Hum. Vaccin. Immunother.}, 8:1694--1702, 2012.

\bibitem{VenslauskasEtAl_SmallMoleculeDeliveryEP_Bioelectrochem2010}
M.~S. Venslauskas, S.~Satkauskas, and R.~Rodaite-Riseviciene.
\newblock Efficiency of the delivery of small charged molecules into cells in
  vitro.
\newblock {\em Bioelectrochem.}, 79:130--135, 2010.

\bibitem{GengAndLu_MicrofluidicEP_LabChip2013}
T.~Geng and C.~Lu.
\newblock Microfluidic electroporation for cellular analysis and delivery.
\newblock {\em Lab Chip}, 13:3803--3821, 2013.

\bibitem{CalvetAndMir_ECTAndImmunotherapy_CancerMetastRev2016}
C.~Y. Calvet and L.~M. Mir.
\newblock The promising alliance of anti-cancer electrochemotherapy with
  immunotherapy.
\newblock {\em Cancer Metastasis Rev.}, 35:165--177, 2013.

\bibitem{BretonMir_MicrosecondNanosecondElectricPulsesCancerTreatments_BEMS2012}
M.~Breton and L.~M. Mir.
\newblock Microsecond and nanosecond electric pulses in cancer treatments.
\newblock {\em Bioelectromagnetics}, 33:106--123, 2012.

\bibitem{JiangEtal_IRE_Review_TBME2015}
C.~Jiang, R.~V. Davalos, and J.~C. Bischof.
\newblock A review of basic to clinical studies of irreversible electroporation
  therapy.
\newblock {\em Trans. Biomed. Eng.}, 62:4--20, 2015.

\bibitem{FrandsenEtal_NormalAndMalignantCellsResponsesCalciumEP_CancerRes2017}
S.~K. Frandsen, M.~B. Kruger, and U.~M. Mangalanathan.
\newblock Normal and malignant cells exhibit differential responses to calcium
  electroporation.
\newblock {\em Cancer Res.}, 2017.
\newblock in press.

\bibitem{Beebe_nsPEFMechanisms_JNanomedRes2015}
S.~J. Beebe.
\newblock Mechanisms of nanosecond pulsed electric field (nspef)-induced cell
  death in cells and tumors.
\newblock {\em J. Nanomed. Res.}, 2:1--5, 2015.

\bibitem{ChopinetAndRols_nsPEF_Review_Bioelectrochem2015}
L.~Chopinet and M.-P. Rols.
\newblock Nanosecond electric pulses: a mini-review of the present state of the
  art.
\newblock {\em Bioelectrochem.}, 103:2--6, 2015.

\bibitem{PakhomovEtAlIbey_CancellationCellularResponses-nsPEF-StimulusPolarityReversal_CellMolLifeSci2014}
A.~G. Pakhomov, I.~Semenov, S.~Xiao, O.~N. Pakhomova, B.~Gregory, K.~H.
  Schoenbach, J.~C. Ullery, H.~T. Beier, S.~R. Rajulapati, and B.~L. Ibey.
\newblock Cancellation of cellular responses to nanoelectroporation by
  reversing the stimulus polarity.
\newblock {\em Cellular Mol. Life Sci.}, 71:4431--4441, 2014.

\bibitem{GianulisEtAlPakhomov_ElectroporationMammalianCellsNanosecondFieldOscillations-FieldReversalInihibition_SciReports2015}
E.~C. Gianulis, J.~Lee, C.~Jiang, S.~Xiao, B.~L. Ibey, and A.~G. Pakhomv.
\newblock Electroporation of mammalian cells by nanosecond electric field
  oscillations and its inhibition by the electric field reversal.
\newblock {\em Sci. Reports}, 5:13818, 2015.

\bibitem{VasilkoskiEtAl_ElectroporationAbsouteRateEquationNanosecondPoreCreation_PhysRevE2006}
Z.~Vasilkoski, A.~T. Esser, T.~R. Gowrishankar, and J.~C. Weaver.
\newblock Membrane electroporation: The absolute rate equation and nanosecond
  timescale pore creation.
\newblock {\em Phys. Rev. {E}}, 74:021904, 2006.

\bibitem{AbidorEtAlChizmadzhevMainFacts1stPaper1979}
I.~G. Abidor, V.~B. Arakelyan, L.~V. Chernomordik, Yu.~A. Chizmadzhev, V.~F.
  Pastushenko, and M.~R. Tarasevich.
\newblock Electric breakdown of bilayer membranes: {I. T}he main experimental
  facts and their qualitative discussion.
\newblock {\em Bioelectrochem. Bioenerget.}, 6:37--52, 1979.

\bibitem{WeaverAndVernier_ComplexDarkPores_arXiv2017}
J.~C. Weaver and P.~T. Vernier.
\newblock Pore lifetimes in cell electroporation: complex dark pores?
\newblock {\em ar{X}iv:1708.07478[physics.bio-ph]}, https://arxiv.org/abs/1708.07478  2017.

\bibitem{SternEtAl_BroadPoreLifetimeDistributions_arXiv2017}
J.~V. Stern, T.~R. Gowrishankar, K.~C. Smith, and J.~C. Weaver.
\newblock Broad pore lifetime distributions: a fundamental concept for cell
  electroporation.
\newblock {\em ar{X}iv:1708.07613[physics.bio-ph]}, https://arxiv.org/abs/1708.07613  2017.

\bibitem{BennettEtAlTieleman_AtomisticSimulationsPoreFormationClosure_BLM_BPJ2014}
W.~F.~D. Bennett, N.~Sapay, and D.~P. Tieleman.
\newblock Atomistic simulations of pore formation and closure in lipid
  bilayers.
\newblock {\em Biophysical J.}, 106:210--219, 2014.

\bibitem{LevineVernier_LifeCyclePoreStepsCreationAnnihilation_JMemBiol2010}
Z.~A. Levine and P.~T. Vernier.
\newblock Life cycle of an electropore: Field-dependent and field-independent
  steps in pore creation and annihilation.
\newblock {\em J. Memb. Biol.}, 236:27--36, 2010.

\bibitem{PakhomovEtAlPakhomova_MultipleNanosecondPulsesIncreaseNumber-NotLongLivedPoreSize_BBA2015}
A.~G. Pakhomov, E.~Gianulis, P.~T. Vernier, I.~Semenov, S.~Xiao, and
  O.~Pakhomova.
\newblock Multiple nanosecond electric pulses increase the number but not the
  size of long-lived nanopores in the cell membrane.
\newblock {\em Biochim. Biophys. Acta}, 1848:958--966, 2015.

\bibitem{CanatellaEtAlPrausnitz_ElectroporationUptakeViabilityQuantitativeStudy_BPJ2001}
P.~J. Canatella, J.~F. Karr, J.~A. Petros, and M.~R. Prausnitz.
\newblock Quantitative study of electroporation-mediated molecular uptake and
  cell viability.
\newblock {\em Biophysical J.}, 80:755--764, 2001.

\bibitem{PakhomovEtAl_nsPEF_LongLastingSinglePulsePermeabilizationPM_BEMS2007}
A.~G. Pakhomov, J.~F. Kolb, J.~A. White, R.~P. Joshi, S.~Ziao, and K.~H.
  Schoenbach.
\newblock Long-lasting membrane permeabilzation in mammalian cells by
  nanosecond pulsed electric field (ns{PEF}).
\newblock {\em Bioelectromagnetics.}, 28:655--663, 2007.

\bibitem{KennedyEtAlBooske_QuantificationElectroporationKineticsPI_FinalVersion_BPJ2008}
S.~M. Kennedy, Z.~Ji, J.~C. Hedstrom, J.~H. Booske, and S.~C. Hagness.
\newblock Quantitation of electroporation uptake kinetics and electric field
  heterogeneity effects in cells.
\newblock {\em Biophys. J.}, 94:5018--5027, 2008.

\bibitem{PakhomovaEtAlPakhomov_Electroporationn-InducedElectrosensitization_PLoS_ONE2011}
O.~N. Pakhomova, B.~W. Gregory, V.~A. Khorokhorina, A.~M. Bowman, S.~Xiao, and
  A.~G. Pakhomov.
\newblock Electroporation-induced electrosensitization.
\newblock {\em {PL}o{S} {ONE}}, 6:e17100, 2011.

\bibitem{SmithEtAl_EmergencelargePoreSubpopulationDuringElectroporationPulses2013}
K.~C. Smith, R.~S. Son, T.~R. Gowrishankar, and J.~C. Weaver.
\newblock Emergence of a large pore subpopulation during electroporating
  pulses.
\newblock {\em Bioelectrochemistry}, 100:3 -- 10, 2014.

\bibitem{SmithWeaver_ActiveMechanismsNeededSubmicrosecondMVperMeterPulses_BPJ2008}
K.~C. Smith and J.~C. Weaver.
\newblock Active mechanisms are needed to describe cell responses to
  submicrosecond, megavolt-per-meter pulses: Cell models for ultrashort pulses.
\newblock {\em Biophys. J.}, 95:1547--1563, 2008.

\bibitem{GowrishankarEtAl_TransportBasedModelsElectricFieldCellResponses_Review_ProcIEEE2012}
T.~R. Gowrishankar, K.~C. Smith, and J.~C. Weaver.
\newblock Transport-based biophysical system models of cells for quantitatively
  describing responses to electric fields.
\newblock {\em Proc {IEEE}}, 101:505--517, 2013.

\bibitem{DeBruinKrassowska_TheoreticalModel_SingleCellEporeI_FieldStrength_RestPotential_BPJ1999}
K.~A. De{B}ruin and W.~Krassowska.
\newblock Modeling electroporation in a single cell: {I}. {E}ffects of field
  strength and rest potential.
\newblock {\em Biophys. J.}, 77:1213--1224, 1999.

\bibitem{JoshiSchoenbach_Review_nsPEF_CritRevBME2010}
R.~P. Joshi and K.~H. Schoenbach.
\newblock Bioelectric effects of intense ultrashort pulses.
\newblock {\em Crit. Rev. Biomed. Engr.}, 38:255--304, 2010.

\bibitem{LiLin_NumericalSimulationMolecularUptakeElectroporation_BChem2011}
J.~Li and H.~Lin.
\newblock Numerical simulation of molecular uptake via electroporation.
\newblock {\em Bioelectrochemistry}, 82:10--21, 2011.

\bibitem{SadikEtAlLin_QuantiativePropidiumIodideExperimentsModel-ms-duration_BBA2013}
M.~M. Sadik, J.~Li, J.~W. Shan, D.~I. Shreiber, and H.~Lin.
\newblock Quantification of propidium iodide delivery using millisecond
  electric pulses: Experiments.
\newblock {\em Biochim. Biophys. Acta}, 1828:1322--1328, 2013.

\bibitem{SengelWallace_ImagingDynamics-IndividualElectropores_PlanarDropletInterfaceBilayers_PNAS2016}
J.~T. Sengel and M.~I. Wallace.
\newblock Imaging the dynamics of individual electropores ({E}pub ahead of
  print).
\newblock {\em Proc. Nat. Acad. Sci.}, 113:5281--5286, 2016.

\bibitem{SengelAndWallace_PotentialEnergyBarrier_BLM_EP_PhilTransRoySocB2017}
J.~T. Sengel and M.~I. Wallace.
\newblock Measuring the potential energy barrier to lipid bilayer
  electroporation.
\newblock {\em Phil. Trans. Roy. Soc. B}, 372:201602227, 2017.

\end{thebibliography}

\pagebreak

\begin{figure}
\begin{center}
\begin{tabular}{c}
\includegraphics[width=5.0in]{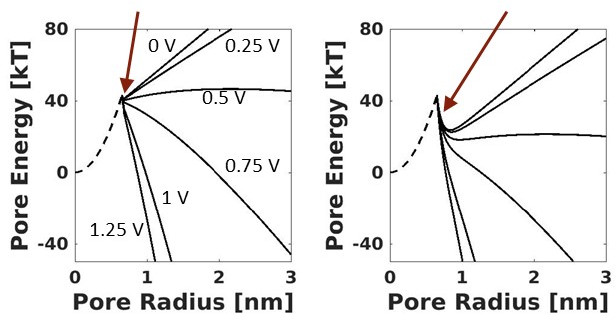}\\ 
\end{tabular}
\end{center}
\caption{\textbf{Pore energy landscapes.} The pore energy landscape
is shown for two different pore lifetimes (100 ns; left) and 
(4 s; right) at different transmembrane
voltages.  As the pore lifetime increases, the energy barrier for 
pore destruction, $\Delta$W, increases reducing the probability of 
overcoming the energy barrier.  The energy barrier for destruction is 
5 kT for 100 ns pore life time and 21 kT for 4 s pore lifetime.
For transmembrane voltages above 0.75 V,
the landscape favors pore expansion.  However, a dramatic increase in
membrane conductance (initiated by pore creation) limits the transmembrane
voltage to less than 0.6 V.
}
\end{figure}

\pagebreak

\begin{figure}
\begin{tabular}{ccc}
\includegraphics[width=2.2in]{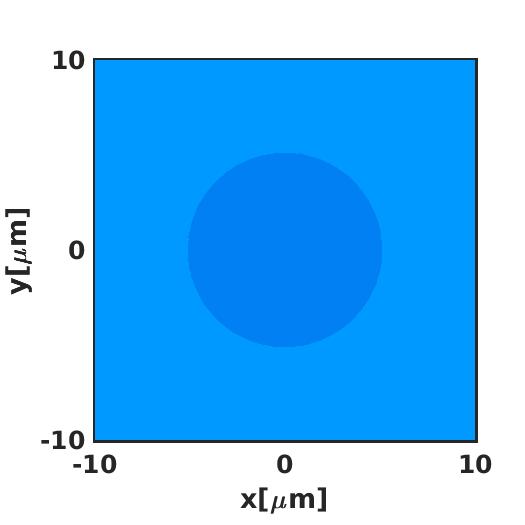} & 
\includegraphics[width=2.2in]{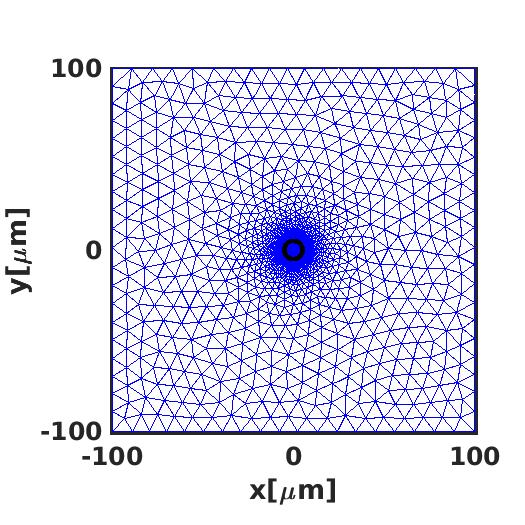} & 
\includegraphics[width=2.2in]{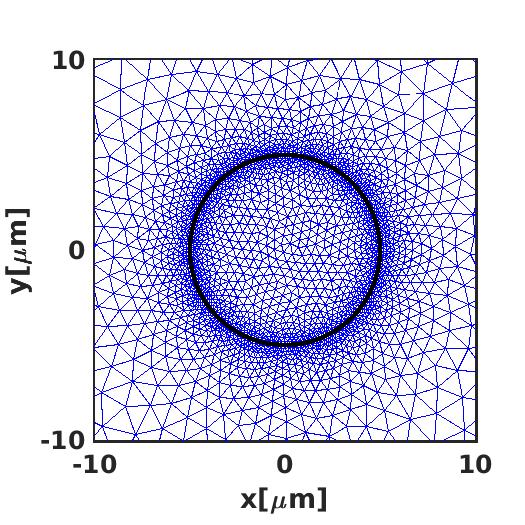} \\ 
\textbf{(a)} & \textbf{(b)} & \textbf{(c)}\\ 
\end{tabular}
\caption{\textbf{Simulation geometry.} The 200 $\mathrm{200\ \mu m\times 200 \ \mu m}$ system model 
contains a $\mathrm{5\ \mu m}$ radius cylindrical cell.  The meshed 
transport network cell model \textit{[REFs Smith Thesis 2011]} is 
represented by 150 membrane node-pairs that describe local transmembrane
voltage, pore distribution, hindrancce, partitioning of solutes and ions
into the pores, and molecular transport.  The 4-nm thick membrane has
a resting potential of -50 mV.  The field is applied between the top and
bottom set of nodes of the simulation box.  Each of the local areas
associated with a transmembrane node-pair is regarded as a very small
planar membrane patch endowed with a resting potential source and a 
complete dynamic EP model.
}
\end{figure}

\pagebreak

\begin{figure}
\begin{tabular}{ccc}
\includegraphics[width=2.2in]{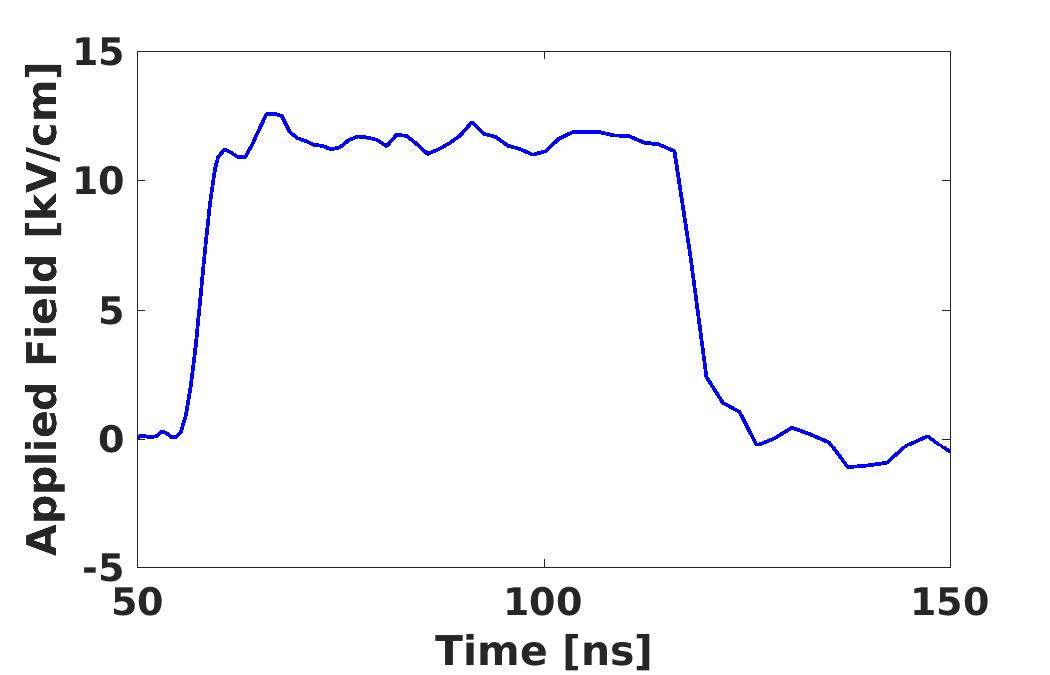} &
\includegraphics[width=2.2in]{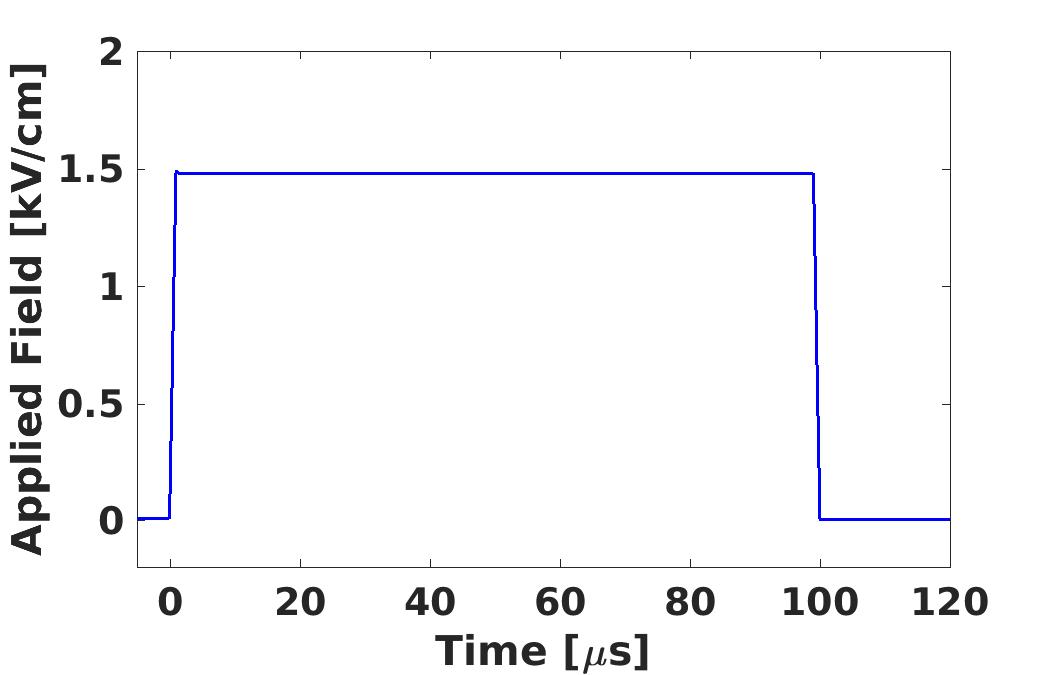} & 
\includegraphics[width=2.2in]{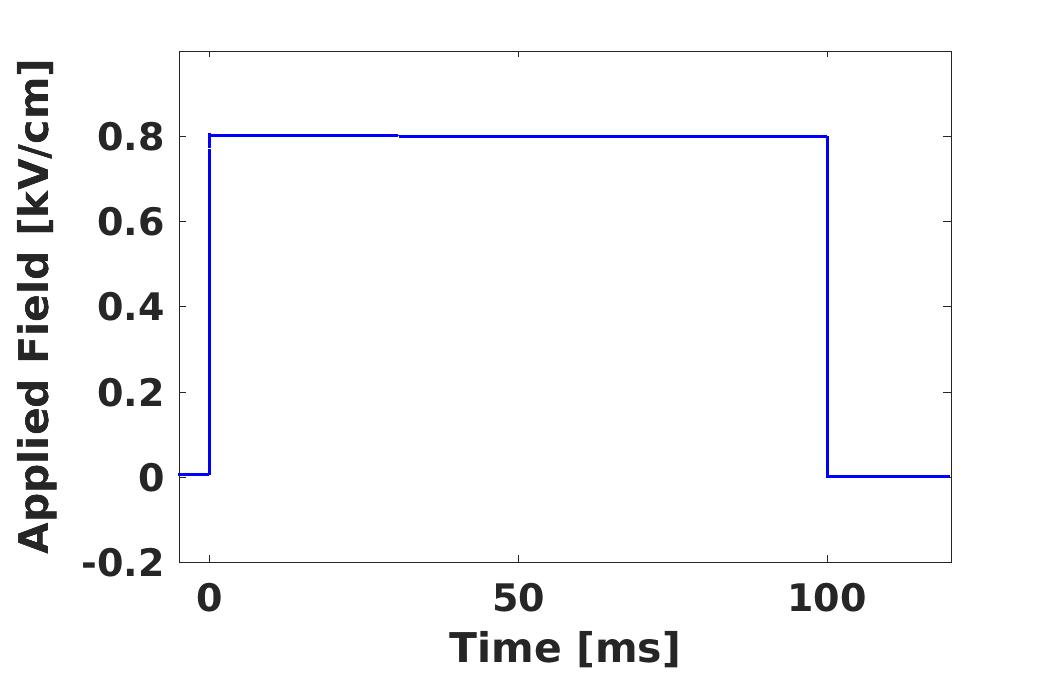} \\
\textbf{(a)} & \textbf{(b)} & \textbf{(c)} \\ 
\end{tabular}
\caption{\textbf{Applied field pulse.}  Three electric field pulses 
are considered: \textbf{(a)} conventional EP pulse
(1.5 kV/cm, \textbf{(b)} 100 $\mathrm{\mu s}$ idealized 
trapezoidal pulse with 1 $\mathrm{\mu s}$ rise and fall times) 
and \textbf{(c)} a nsPEF experimental pulse (13.3 kV/cm, 60 ns)
\cite{PakhomovaEtAlPakhomov_Electroporationn-InducedElectrosensitization_PLoS_ONE2011}.
  The cell model of Fig. 1 are subject to these two
field pulses.  The local distribution of transmembrane voltage, membrane
conductance and pore distribution are dependent on each other.  The coupled
system is solved in Matlab (Mathworks, Natick, MA).
}
\end{figure}

\pagebreak

\begin{figure}
\begin{tabular}{ccc}
\includegraphics[width=2.2in]{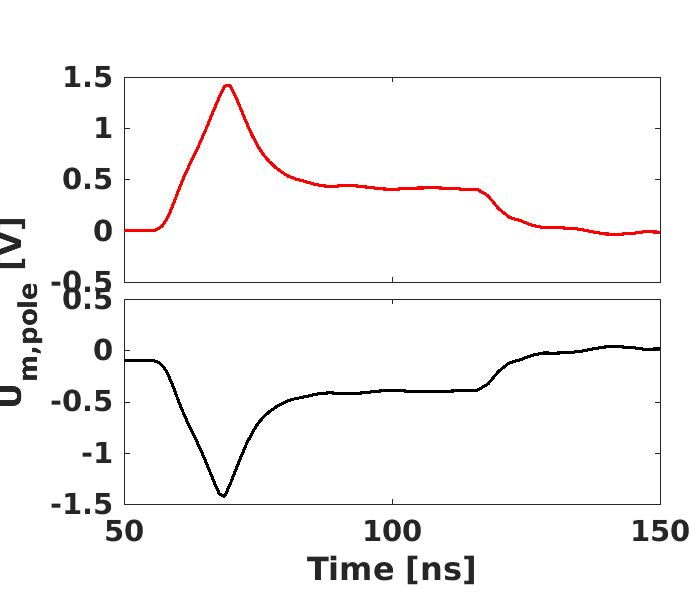} & 
\includegraphics[width=2.2in]{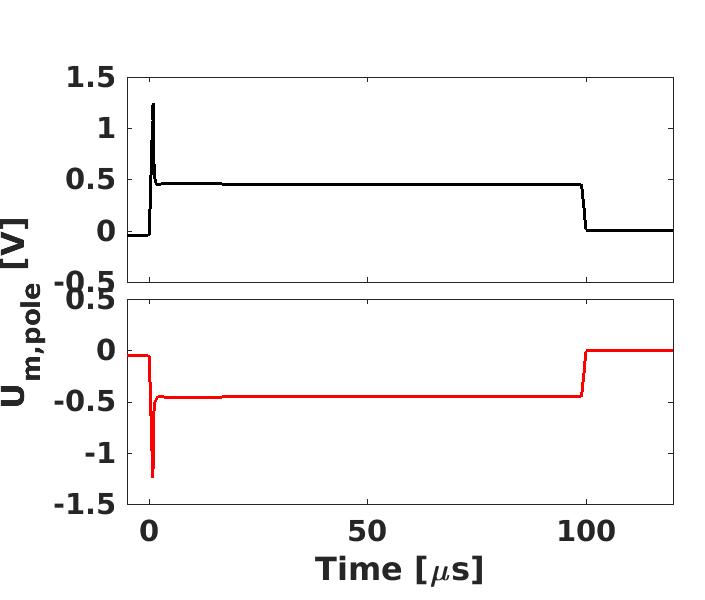} & 
\includegraphics[width=2.2in]{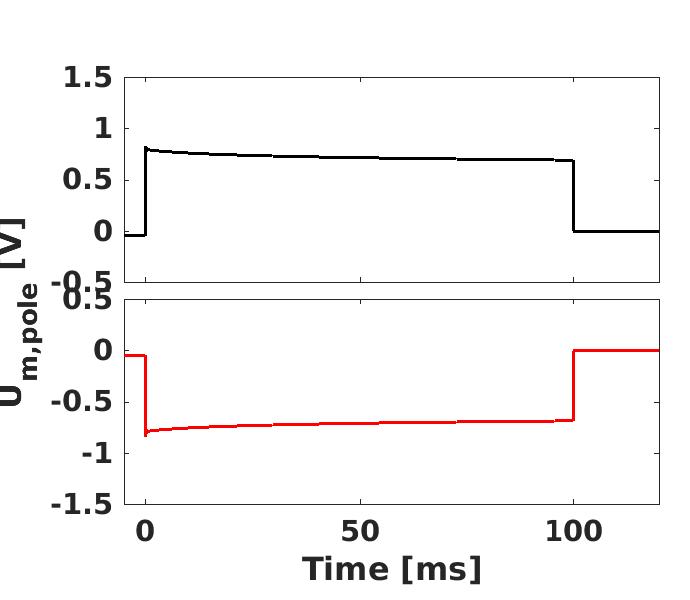} \\ 
\textbf{(a)} & \textbf{(b)} & \textbf{(c)} \\ 
\end{tabular}
\caption{\textbf{Transmembrane voltage response.}  Both nsPEF \textbf{(a)} 
and conventional EP pulses \textbf{(b)} cause reversible electrical
breakdown (REB) leading to a $\mathrm{U_m}$ plateau of 0.5 V.  However,
the 100 ms pulse \textbf{(c)} does not exhibit REB, but show a decline
in transmembrane voltage during the pulse.   The response
is identical for all three pore lifetimes for each pulse as the pore 
lifetime primarily affects the destruction of pores after the pulse.
}
\end{figure}

\pagebreak

\begin{figure}
\begin{tabular}{ccc}
\includegraphics[width=2.2in]{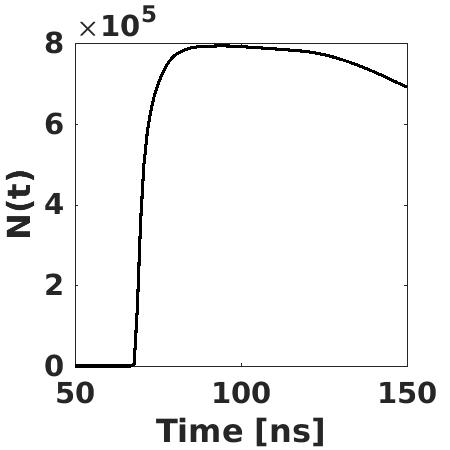} & 
\includegraphics[width=2.2in]{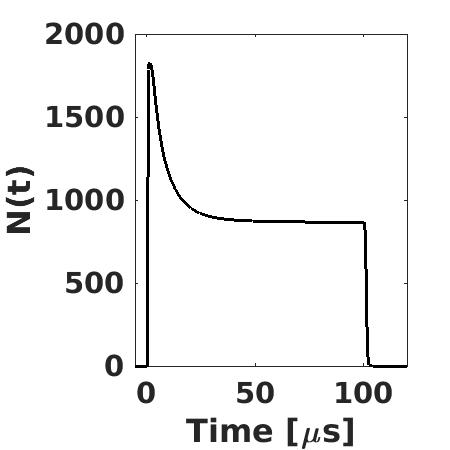} & 
\includegraphics[width=2.2in]{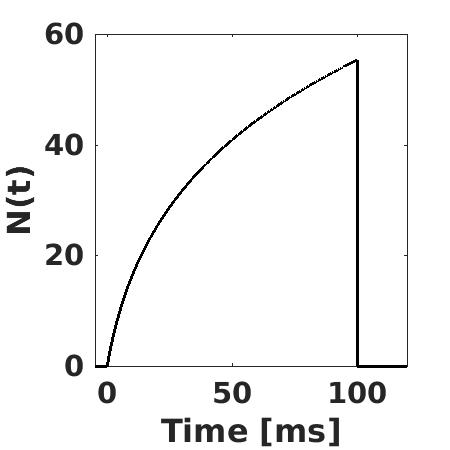} \\ 
\textbf{(a)} & \textbf{(b)} & \textbf{(c)} \\ 
\includegraphics[width=2.2in]{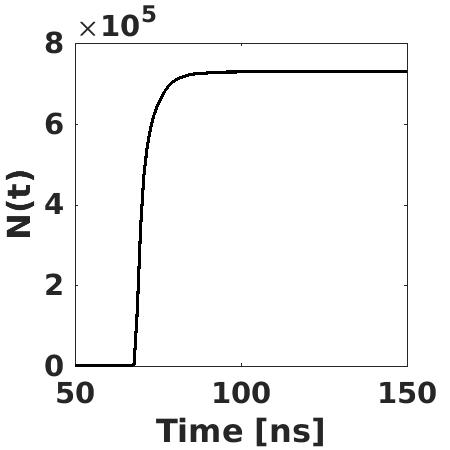} & 
\includegraphics[width=2.2in]{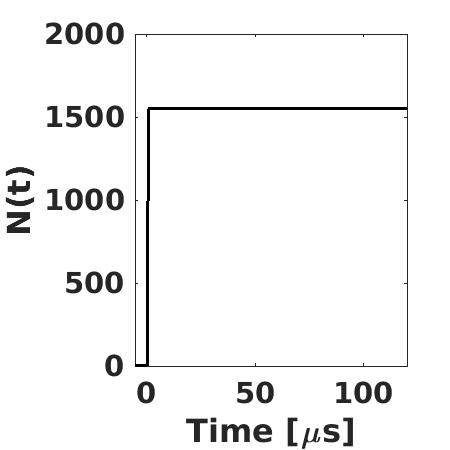} & 
\includegraphics[width=2.2in]{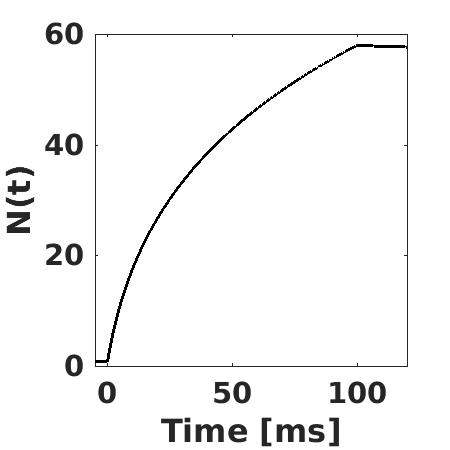} \\
\textbf{(d)} & \textbf{(e)} & \textbf{(f)} \\ 
\end{tabular}
\caption{\textbf{Pore number response.}  Top row shows the pore number 
response during the pulse for 
100 ns pore lifetime and the bottom row shows the response for a 4 s pore
lifetime. Response is shown for three pulses: 13.3 kV/cm, 60 ns nsPEF pulse 
[\textbf{(a),(d)}], 1.5 kV/cm, 100 $\mathrm{\mu}$s conventional EP pulse
[\textbf{(b), (e)}], and a 0.8 kV/cm, 100 ms long EP pulse 
[\textbf{(c), (f)}].  Although the nsPEF pulse 
is short, pore loss begins during the pulse for 100 ns pore lifetime 
\textbf{(a)}.  
However, for a conventional pulse, there
is a significant decrease in the number of pores even during the pulse.
\textbf{(b)}.  
For the long pulse, the pore numbers are small because the applied field
is low (0.8 kV/cm) and the pore number does not decline during the pulse
\textbf{(c)}.
For a  4 spore lifetime, the pore number does not
decline during the pulse \textbf{(d-f)}.  
Because a nsPEF pulse leads to supra-EP, the number of pores
is two orders of magnitude bigger than that for a conventional pulse
\textbf{(d,e)}.  
}
\end{figure}

\pagebreak

\begin{figure}
\begin{tabular}{cccc}
\includegraphics[width=1.6in]{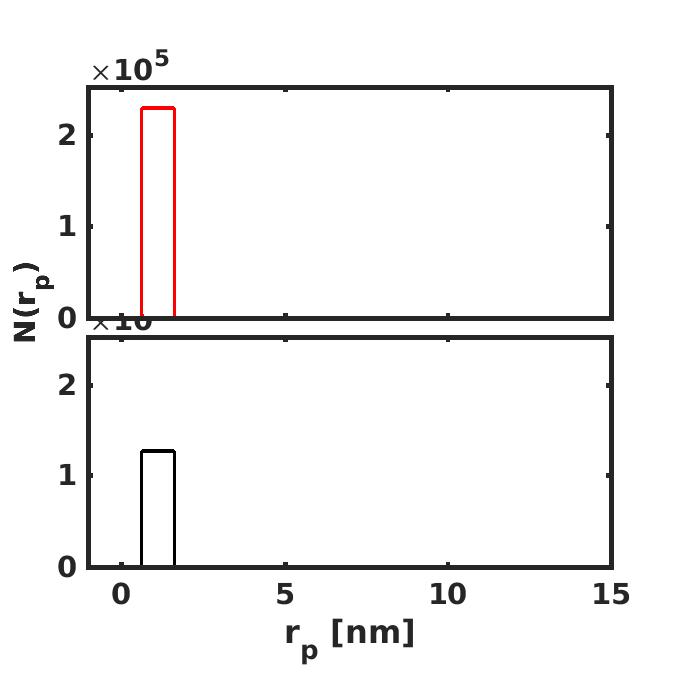} & 
\includegraphics[width=1.6in]{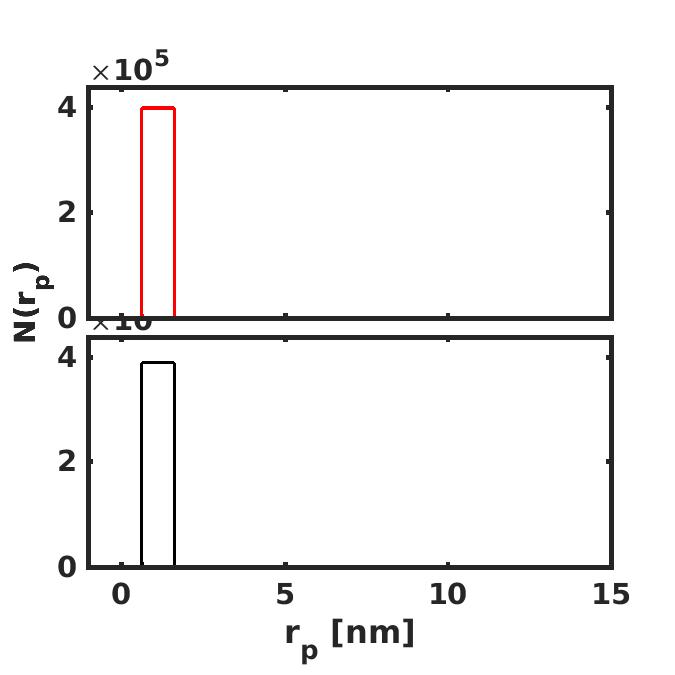} & 
\includegraphics[width=1.6in]{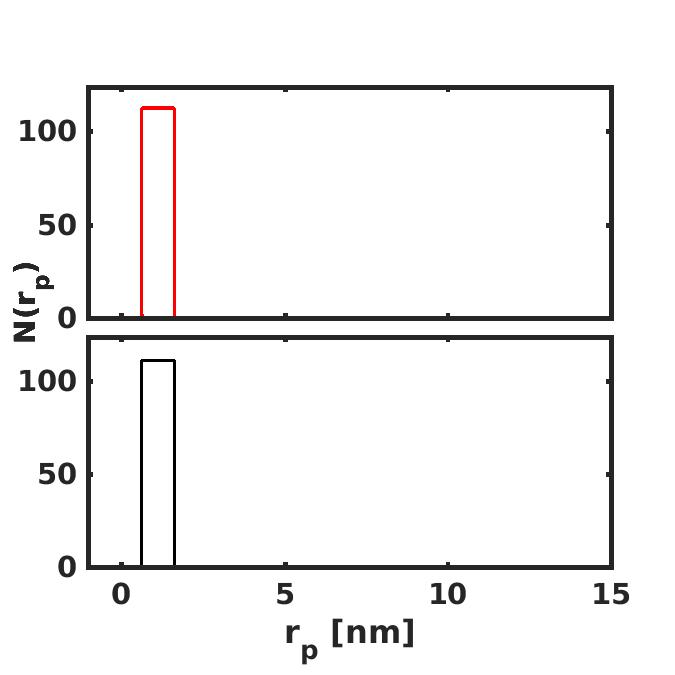} & 
\includegraphics[width=1.6in]{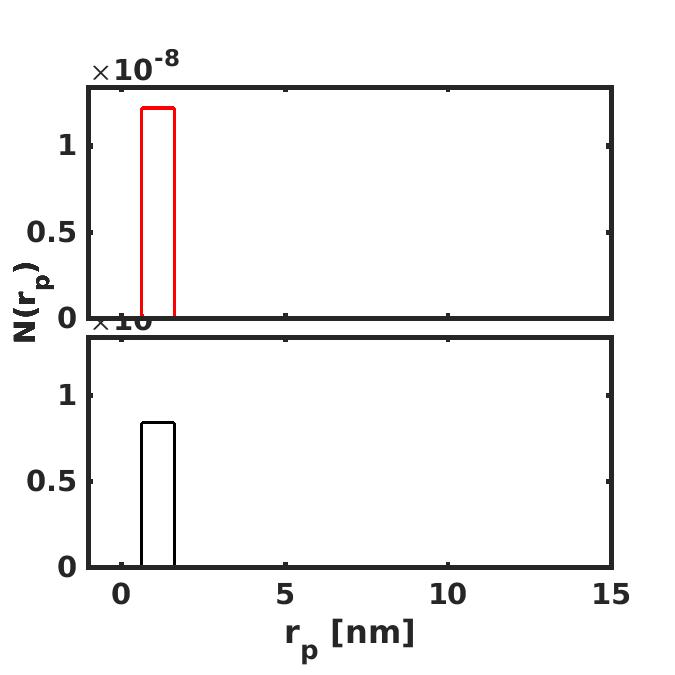} \\ 
\textbf{(a)} & \textbf{(b)} & \textbf{(c)} & \textbf{(d)} \\ 
\includegraphics[width=1.6in]{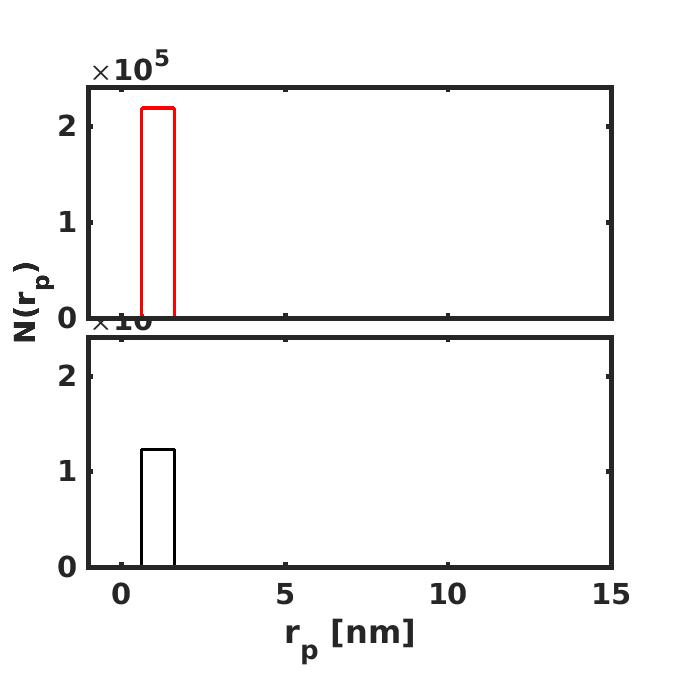} & 
\includegraphics[width=1.6in]{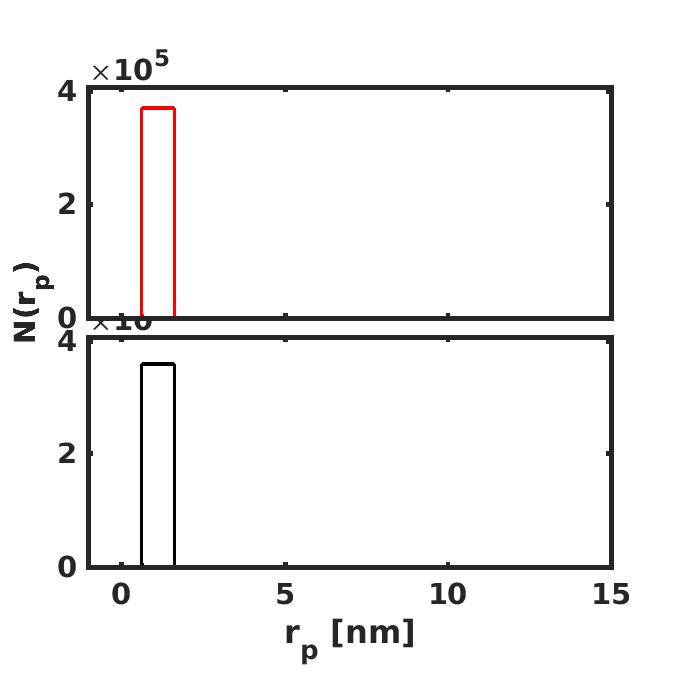} & 
\includegraphics[width=1.6in]{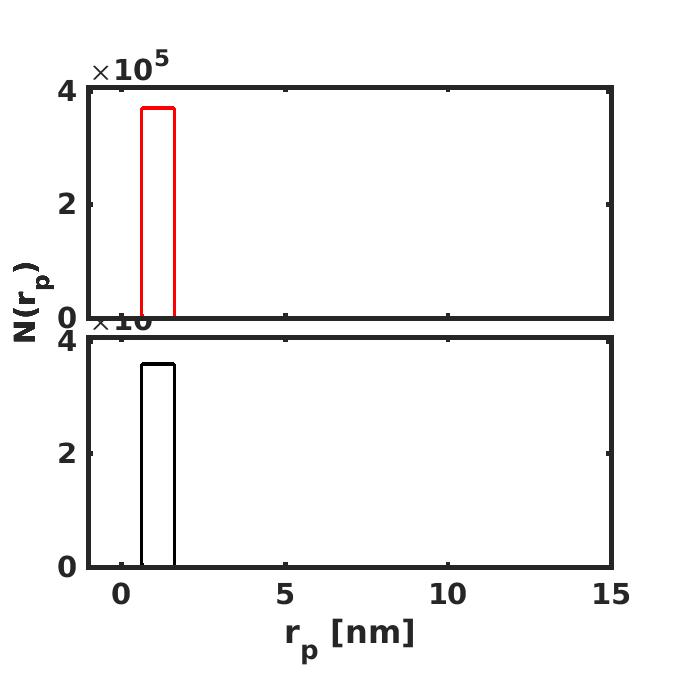} & 
\includegraphics[width=1.6in]{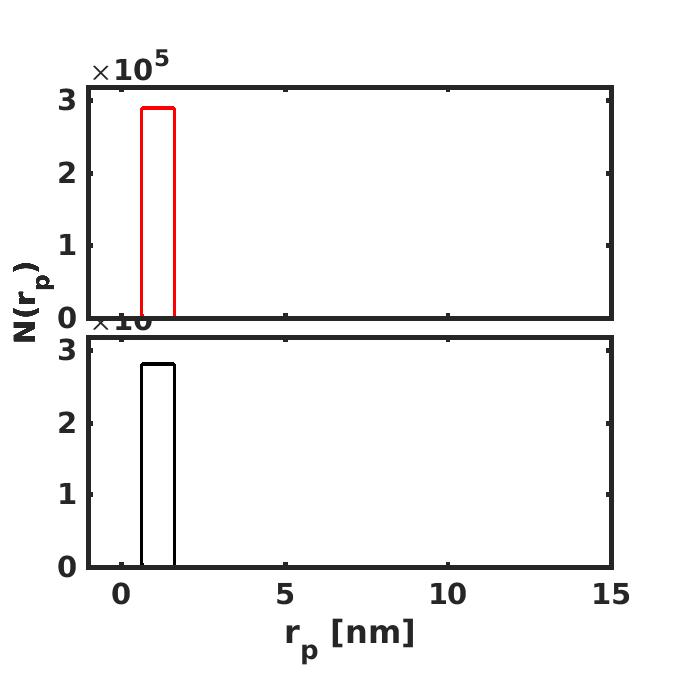} \\ 
\textbf{(e)} & \textbf{(f)} & \textbf{(g)} & \textbf{(h)} \\ 
\end{tabular}
\caption{\textbf{Pore histogram for the nsPEF (13.3 kV/cm, 60 ns) pulse.}  
Pore histograms are shown at the start of pulse (\textbf{a, e}, 10 ns), 
end of pulse (\textbf{b, f}, 60 ns),
shortly after the pulse (\textbf{c, g}, 1 $\mathrm{mu}$s), 
and long after the pulse (\textbf{d, h}, 1 s) for
100 ns pore lifetime (top row) and 4 s pore lifetime (bottom row).  
Because of the large amplitude of the applied field, a large numbmer of 
pores are created.  However, the pores do not have sufficient time to 
expand during the short pulse.
}
\end{figure}

\pagebreak

\begin{figure}
\begin{tabular}{cccc}
\includegraphics[width=1.6in]{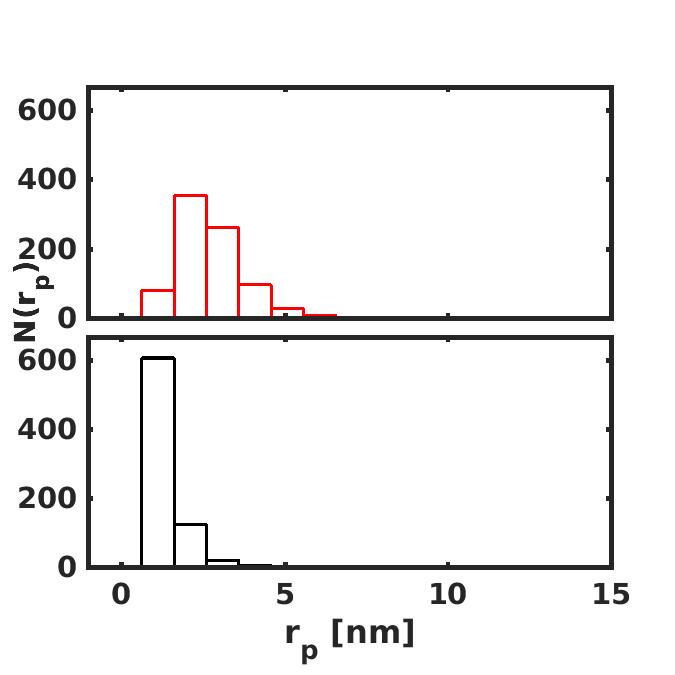} & 
\includegraphics[width=1.6in]{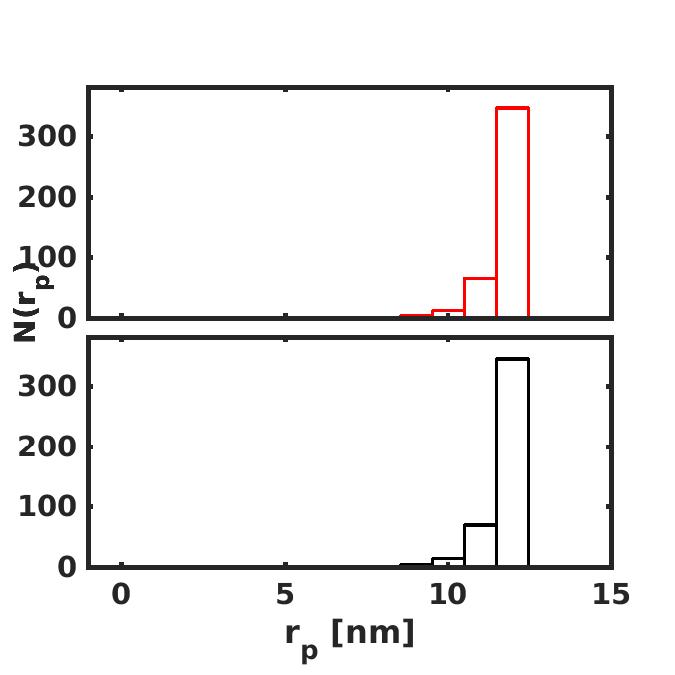} & 
\includegraphics[width=1.6in]{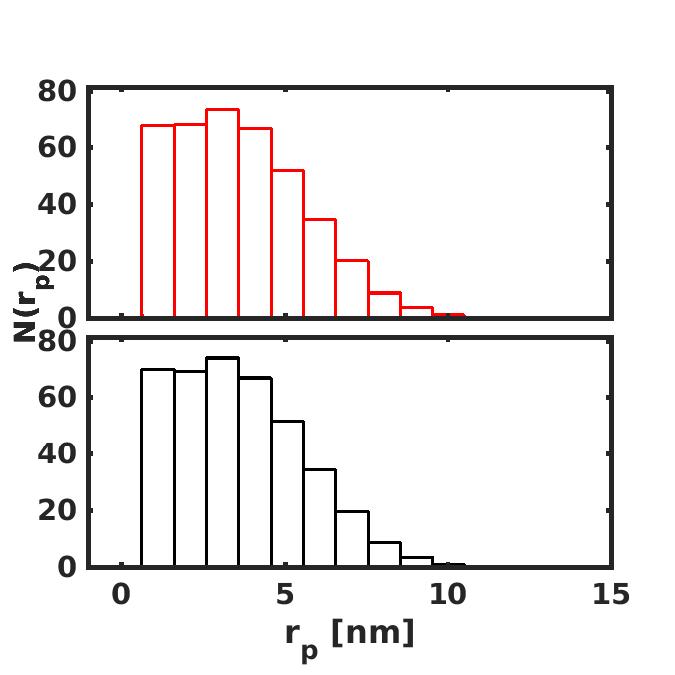} & 
\includegraphics[width=1.6in]{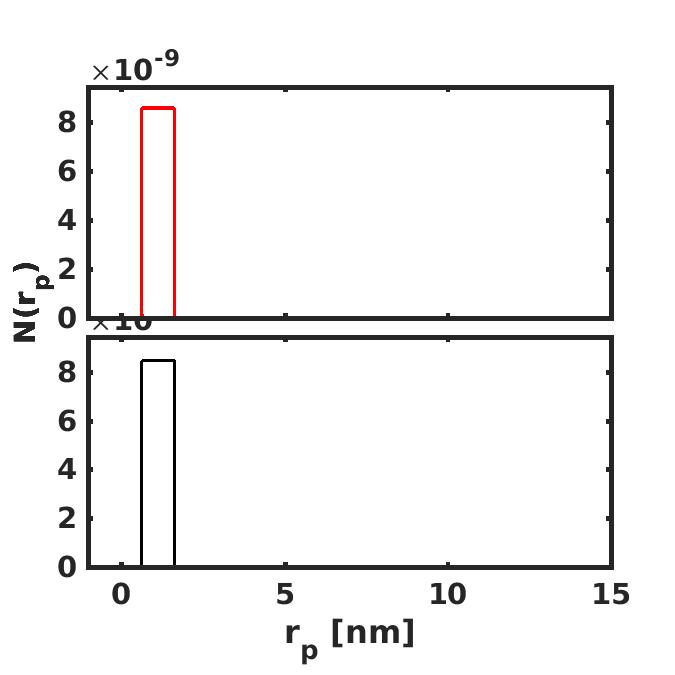} \\ 
\textbf{(a)} & \textbf{(b)} & \textbf{(c)} & \textbf{(d)} \\ 
\includegraphics[width=1.6in]{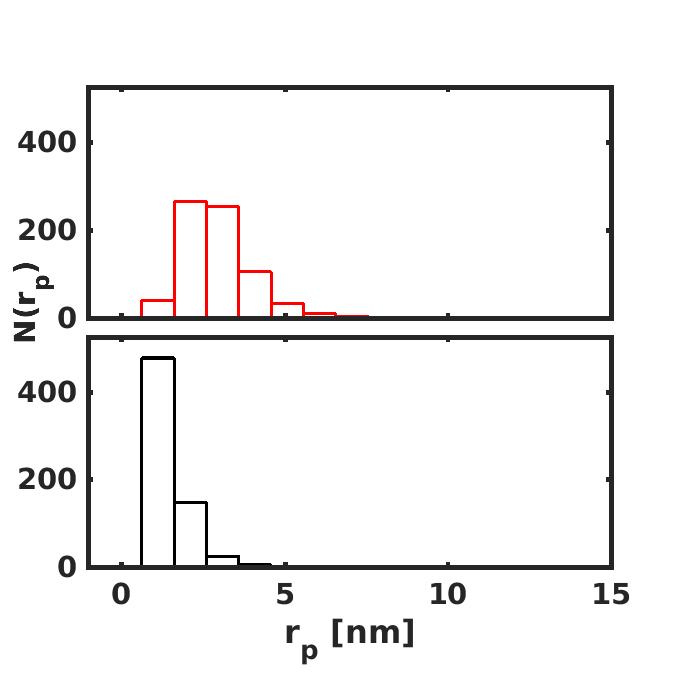} & 
\includegraphics[width=1.6in]{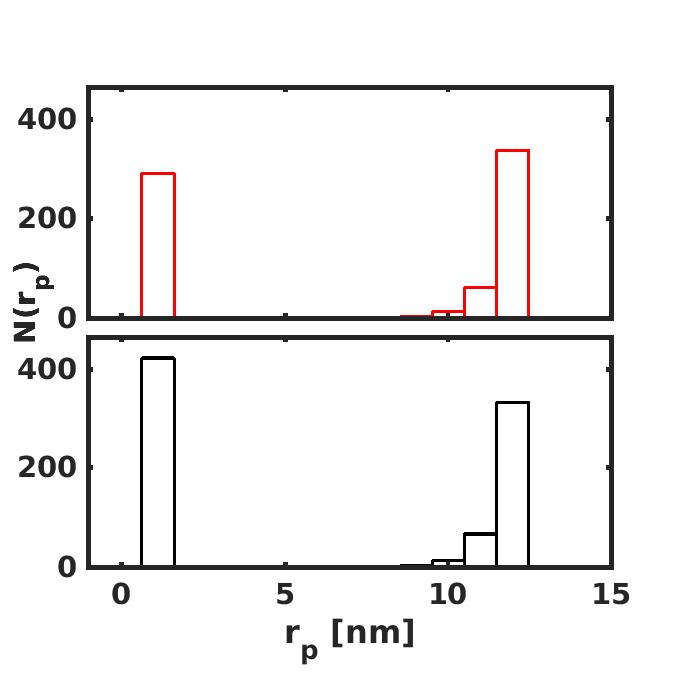} & 
\includegraphics[width=1.6in]{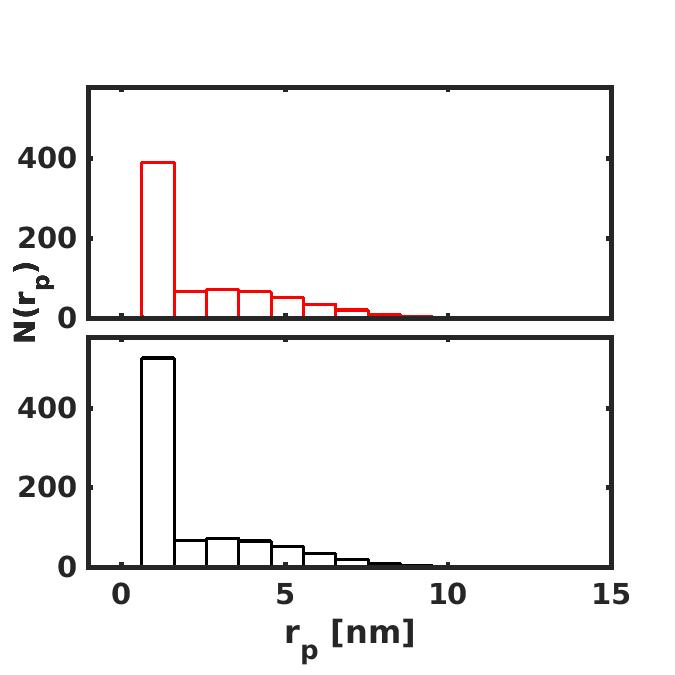} & 
\includegraphics[width=1.6in]{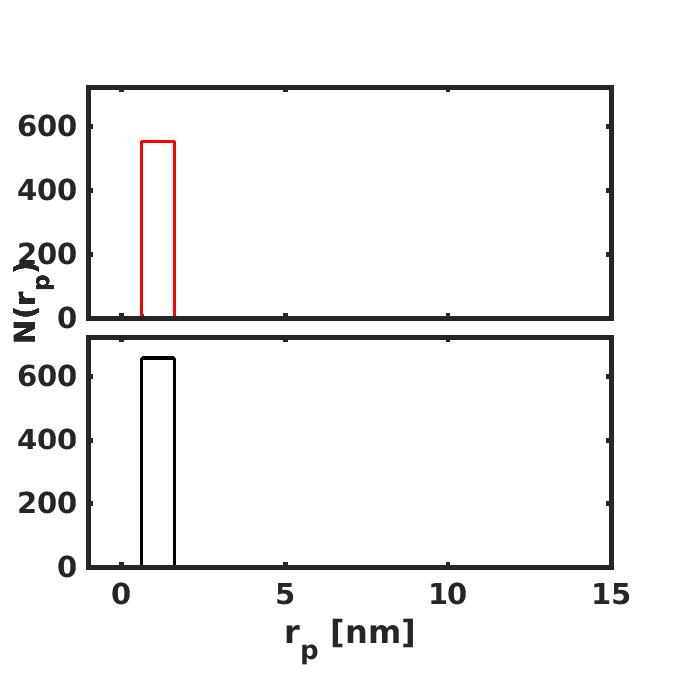} \\ 
\textbf{(e)} & \textbf{(f)} & \textbf{(g)} & \textbf{(h)} \\ 
\end{tabular}
\caption{\textbf{Pore histogram for the conventional EP (1.5 kV/cm, 100 $\mathrm{\mu}$s) pulse.}  
Pore histograms are shown at the start of pulse (\textbf{a, e}, 1 $\mathrm{\mu}$s), 
end of pulse (\textbf{b, f}, 99 $\mathrm{\mu}$s),
shortly after the pulse (\textbf{c, g}, 101 $\mathrm{\mu}$s), 
and long after the pulse (\textbf{d, h}, 1 s) for
100 ns pore lifetime (top row) and 4 s pore lifetime (bottom row).  
The conventional pulses last long enough to expand the pores to 12 nm.
But, these pores begin to shrink within 1 $\mathrm{\mu}$s after the
pulse ends.  For the short pore lifetime, most pores disappear by 1 s.  However,
for longer pore lifetimes, a significant number of minimum-sized pores
persist long after the pulse.
}
\end{figure}

\pagebreak

\begin{figure}
\begin{tabular}{cccc}
\includegraphics[width=1.6in]{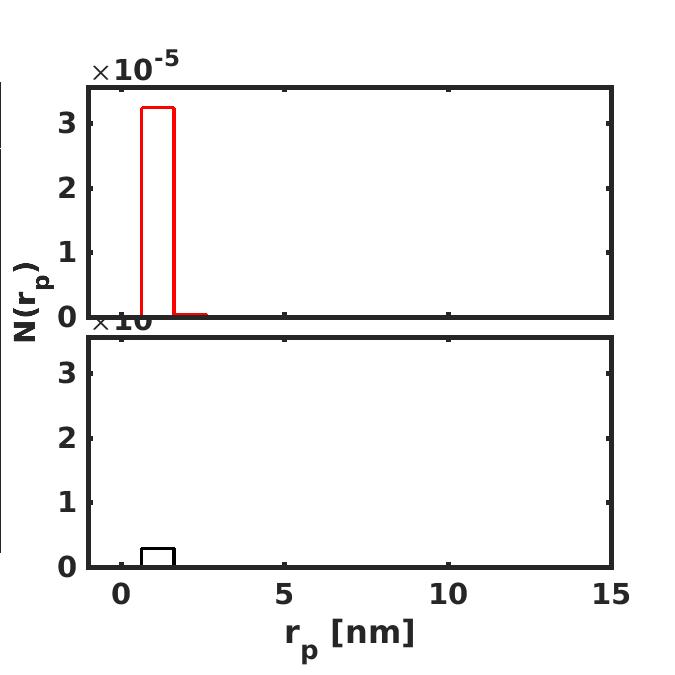} & 
\includegraphics[width=1.6in]{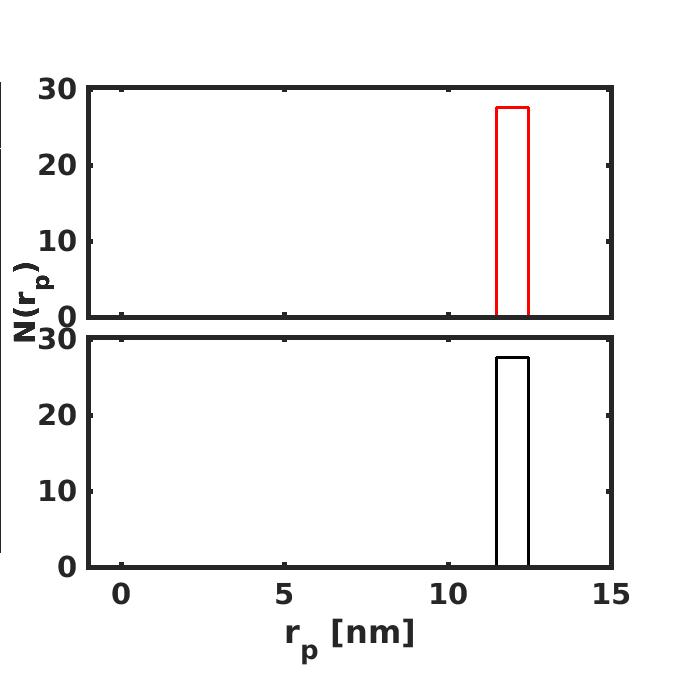} & 
\includegraphics[width=1.6in]{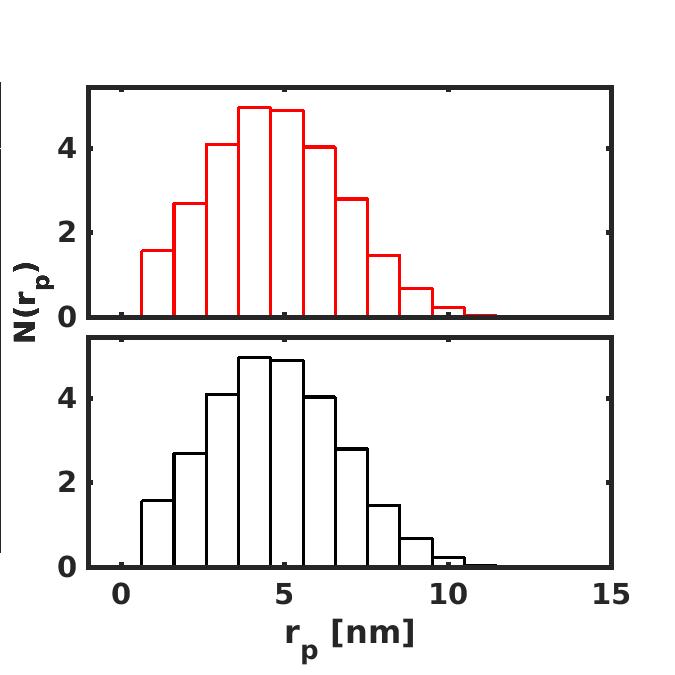} & 
\includegraphics[width=1.6in]{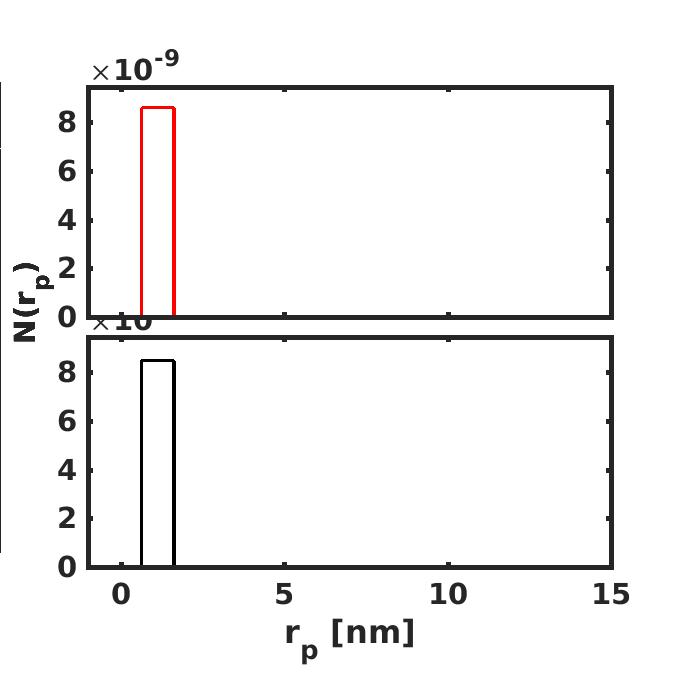} \\ 
\textbf{(a)} & \textbf{(b)} & \textbf{(c)} & \textbf{(d)} \\ 
\includegraphics[width=1.6in]{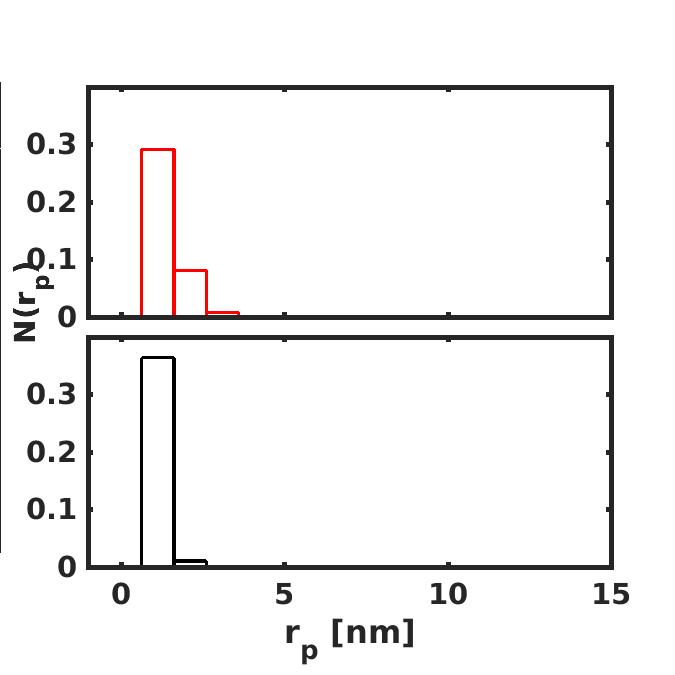} & 
\includegraphics[width=1.6in]{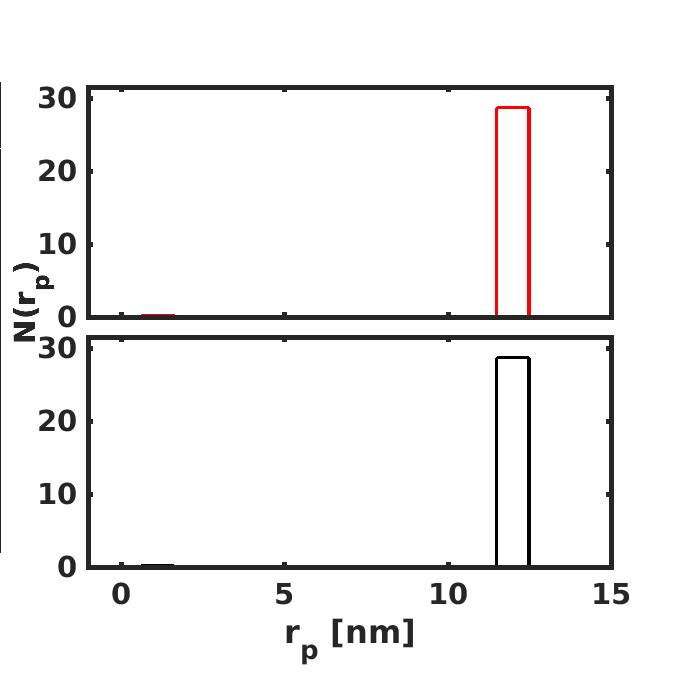} & 
\includegraphics[width=1.6in]{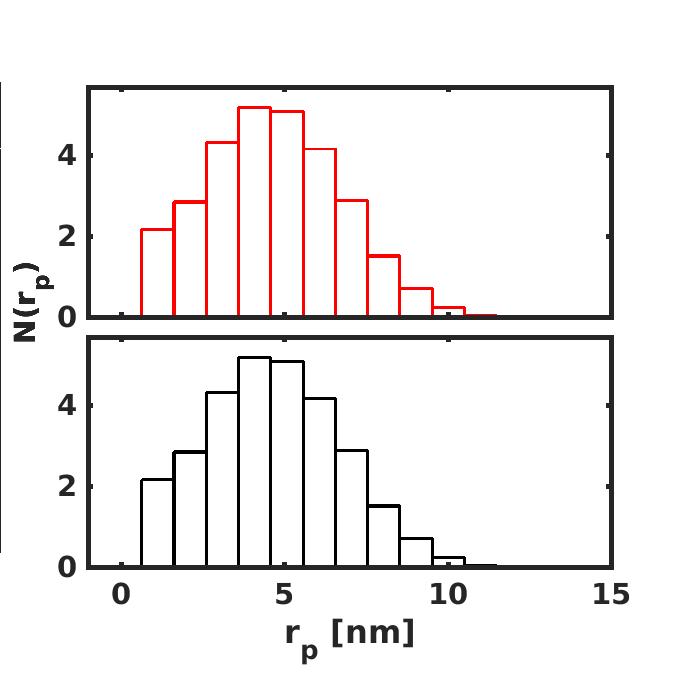} & 
\includegraphics[width=1.6in]{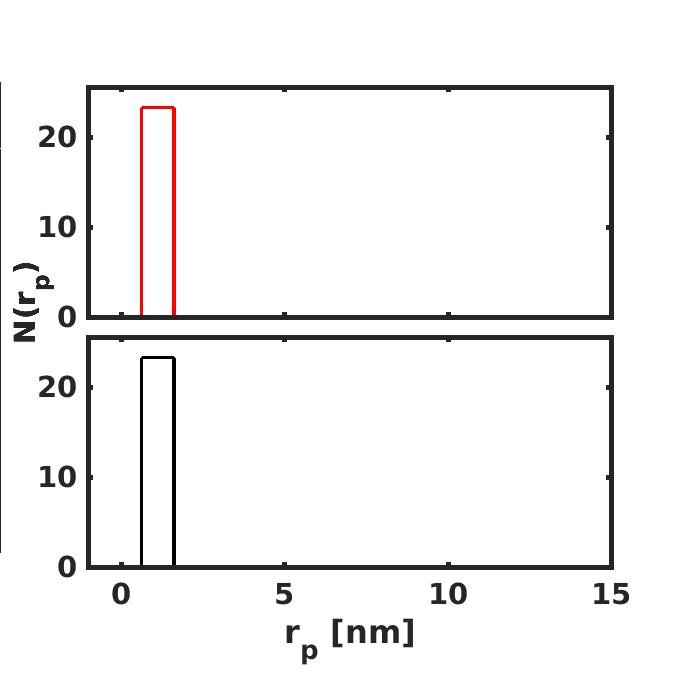} \\ 
\textbf{(e)} & \textbf{(f)} & \textbf{(g)} & \textbf{(h)} \\ 
\end{tabular}
\caption{\textbf{Pore histogram for the long (0.8 kV/cm, 100 ms) pulse.}  
Pore histograms are shown at the start of pulse (\textbf{a, e}, 1 $\mathrm{\mu}$s), 
end of pulse (\textbf{b, f}, 99 ms),
shortly after the pulse (\textbf{c, g}, 100.001 ms), 
and long after the pulse (\textbf{d, h}, 1 s) for
100 ns pore lifetime (top row) and 4 s pore lifetime (bottom row).  
The conventional pulses last long enough to expand the pores to 12 nm.
But, these pores begin to shrink within 1 $\mathrm{\mu}$s after the
pulse ends.  For the short pore lifetime, most pores disappear by 1 s.  However,
for longer pore lifetimes, a significant number of minimum-sized pores
persist long after the pulse.
}
\end{figure}

\clearpage

\begin{table}
\begin{center}
\begin{tabular}{ c l l} \hline
\textbf{Symbol} & \textbf{Description} & \textbf{Value} \\ \hline
r$_*$ & Pore radius at local energy maximum & 0.65 nm \\
B & Steric repulsion constant  & 1.26$\times$10$^{-19}$ J\\
b & Steric repulsion constant & 8.65\\
C & Steric repulsion constant & -2.21$\times$10$^{-20}$ J\\
$\gamma$ & Pore line tension & 2$\times$10$^{-11}$ J/m\\
$\Gamma$ & Membrane tension  &  1$\times$10$^{-5}$ J/m$^2$\\
$F_{max}$ & Membrane electric force for $\Delta\phi_m$=1 V & 6.9$\times$10$^{-10}$ N/V$^2$\\
$r_h$ & Electric force constant & 0.95 nm\\
$r_t$ & Electric force constant & 0.23 nm\\
\end{tabular}
\end{center}
\caption{Simulation parameters}
\end{table}

\end{document}